\documentclass[12pt,a4paper]{article}

\usepackage{amsmath}
\usepackage{amssymb}
\usepackage{epsfig}
\usepackage{graphicx}
\usepackage{cite}

\newcommand{\capdef}{}
\newcommand{\mycaption}[2][\capdef]{\renewcommand{\capdef}{#2}%
       \caption[#1]{{\footnotesize #2}}}
\makeatletter
\renewcommand{\fnum@table}{\textbf{\tablename~\thetable}}
\renewcommand{\fnum@figure}{\textbf{\figurename~\thefigure}}
\makeatother

\newcounter{myenumi}

\renewcommand{\themyenumi}{\roman{myenumi}}
{\end{list}}

\newlength{\myem}
\newcounter{mysubequation}[equation]

\makeatletter
\renewcommand{\section}{\@startsection{section}{1}{0em}{-\baselineskip}%
{\baselineskip}{\normalfont\large\bfseries}}
\renewcommand{\subsection}%
{\@startsection{subsection}{2}{0em}{-0.7\baselineskip}%
{0.7\baselineskip}{\normalfont\bfseries}}
\makeatother

\def\be{\begin{equation}}
\def\ee{\end{equation}}
\newcommand{\beq}{\begin{equation}}
\newcommand{\eeq}{\end{equation}}
\newcommand{\ba}{\begin{array}{c}}
\newcommand{\baz}{\begin{array}{cc}}
\newcommand{\bad}{\begin{array}{ccc}}
\newcommand{\bav}{\begin{array}{cccc}}
\newcommand{\ea}{\end{array}}
\newcommand{\bea}{\begin{equation} \begin{array}{c}}
\newcommand{\eea}{ \end{array} \end{equation}}
\newcommand{\pmns}{\mbox{$ U_{\rm PMNS}$}}
\newcommand{\deltaatm}{\mbox{$\Delta m^2_{31}$}}
\newcommand{\deltasol}{\mbox{$ \Delta m^2_{21}$}}
\newcommand{\dmsol}{\mbox{$\Delta m^2_{\odot}$~}}

\newcommand{\dma}{\Delta m_{31}^2}
\newcommand{\dms}{\Delta m_{21}^2}

\begin{document}


\renewcommand{\thefootnote}{\alph{footnote}}

\begin{flushright}
SISSA 40/2006/EP\\
\end{flushright}

\vspace*{1cm}

\renewcommand{\thefootnote}{\fnsymbol{footnote}}
\setcounter{footnote}{-1}

{\begin{center}
{\Large\textbf{%
Precision Measurement of Solar Neutrino Oscillation\\
Parameters by a Long-Baseline Reactor Neutrino\\
Experiment in Europe\\}}
\end{center}}

\renewcommand{\thefootnote}{\it\alph{footnote}}

\vspace*{.8cm}

{\begin{center}{{\bf
                S.\ T.\ Petcov\footnote[1]{Also at: Institute of 
                Nuclear Research and Nuclear Energy,
                Bulgarian Academy of Sciences, 1784 Sofia, Bulgaria.}
and
                T.\ Schwetz}}
 \end{center}}
{\it
 \begin{center}
  Scuola Internazionale Superiore di Studi Avanzati, and INFN, 
  Sezione di Trieste,\\
  Via Beirut 2--4, I--34014 Trieste, Italy
\end{center}}

\vspace*{0.5cm}

\begin{abstract}
  We consider the determination of the solar neutrino oscillation
  parameters $\dms$ and $\theta_{12}$ by studying oscillations of
  reactor anti-neutrinos emitted by nuclear power plants (located
  mainly in France) with a detector installed in the Frejus underground
  laboratory.  The performances of a water \v{C}erenkov detector of
  147~kt fiducial mass doped with 0.1\% of Gadolinium (MEMPHYS-Gd) and
  of a 50~kt scale liquid scintillator detector (LENA) are compared. In
  both cases 3$\sigma$ uncertainties below 3\% on $\dms$ and of about
  20\% on $\sin^2\theta_{12}$ can be obtained after one year of data
  taking.  The Gadolinium doped Super-Kamiokande detector (SK-Gd) in
  Japan can reach a similar precision if the SK/MEMPHYS fiducial mass
  ratio of 1 to 7 is compensated by a longer SK-Gd data taking time.
  Several years of reactor neutrino data collected by MEMPHYS-Gd or
  LENA would allow a determination of $\dms$ and $\sin^2\theta_{12}$
  with uncertainties of approximately 1\% and 10\% at 3$\sigma$,
  respectively. These accuracies are comparable to those that can be
  reached in the measurement of the atmospheric neutrino oscillation
  parameters $\dma$ and $\sin^2\theta_{23}$ in long-baseline superbeam
  experiments.
\end{abstract}

\newpage

\renewcommand{\thefootnote}{\arabic{footnote}}
\setcounter{footnote}{0}


\section{Introduction}
\indent 

The experiments with solar~\cite{solar,SKsolar,sno}, 
atmospheric~\cite{SKatm}, reactor~\cite{kamland1,kamland}
and accelerator neutrinos~\cite{k2k,MINOS}
have provided during the last several years 
compelling evidence for existence 
of neutrino oscillations caused by nonzero 
neutrino masses and neutrino mixing. 
The data imply the presence of 3-$\nu$ mixing
in the weak charged lepton current 
(see, e.g.,~\cite{STPNu04}): 
\begin{equation}
\nu_{l \mathrm{L}}  = \sum_{j=1}^{3} U_{l j} \, \nu_{j \mathrm{L}},~~
l  = e,\mu,\tau,
\label{3numix}
\end{equation}
%
where $\nu_{l\mathrm{L}}$ are the flavour neutrino fields, 
$\nu_{j \mathrm{L}}$ is the field of neutrino 
$\nu_j$ having a mass $m_j$ and $U$ is the
Pontecorvo--Maki--Nakagawa--Sakata (PMNS) mixing 
matrix~\cite{BPont57}, $U \equiv \pmns$. 
All currently existing $\nu$-oscillation data, except
the data of the LSND experiment~\cite{LSND},
can be described perfectly well 
assuming  3-$\nu$ mixing in vacuum
and we will consider this possibility in what follows.
\footnote{In the LSND experiment indications for $\bar
\nu_{\mu}\to\bar \nu_{e}$ oscillations with $\Delta
m^2_\mathrm{LSND}\simeq 1~\rm{eV}^{2}$ were obtained. The minimal
4-$\nu$ mixing scheme which could incorporate the LSND indications for
$\nu$-oscillations is strongly disfavored by the
data~\cite{Maltoni:2002xd}.  A $\nu$-oscillation explanation of the
LSND results might be possible assuming 5-$\nu$ mixing~\cite{JConrad}.
The LSND results are being tested in the MiniBooNE
experiment~\cite{MiniB}.}

The PMNS matrix can be parametrized by three angles and, depending on
whether the massive neutrinos $\nu_j$ are Dirac or Majorana particles,
by one or three CP-violation (CPV) phases (see, e.g., \cite{BPP1}):
\beq
\pmns = V(\theta_{12},\theta_{13},\theta_{23},\delta)
~{\rm diag}(1, e^{i \frac{\alpha}{2}}, e^{i \frac{\beta}{2}})
\label{pmns}
\eeq
%
where $V(\theta_{12},\theta_{13},\theta_{23},\delta)$ is a CKM-like
matrix, $\delta$ is the Dirac CP-violating phase and $\alpha,\beta$
are two Majorana CPV phases~\cite{BHP80,Doi81SchValle80}.  If we
standardly identify $\dmsol = \Delta m^2_{21} \equiv m^2_2 - m^2_1 >
0$, where $\dmsol$ drives the solar neutrino oscillations, then
$\Delta m^2_{\rm A} = \Delta m^2_{31}\cong \Delta m^2_{32}$,
$\theta_{23} = \theta_{\rm A}$ and $\theta_{12} = \theta_{\odot}$,
$\Delta m^2_{\rm A}$, $\theta_{\rm A}$ and $\theta_{\odot}$ being the
$\nu$-mass squared difference and mixing angles responsible
respectively for atmospheric and solar neutrino oscillations, while
$\theta_{13}$ is the mixing angle constrained by the CHOOZ
experiment~\cite{chooz}.  
The existing neutrino oscillation data allow us to determine $\Delta
m^2_{21}$, $|\Delta m^2_{31}|$, $\sin^2\theta_{12}$ and
$\sin^22\theta_{23}$ with a relatively good precision and to obtain
rather stringent limits on $\sin^2\theta_{13}$ (see,
e.g.,~\cite{BCGPRKL2,TSchwatm05,TSchwSNOW06}).  The best fit values
and the 95\%~C.L.\ allowed ranges of $\Delta m^2_{21}$,
$\sin^2\theta_{12}$, $|\Delta m^2_{31}|$ and $\sin^22\theta_{23}$
read:
\beq
\label{bfvsol}
\begin{array}{ll}
\deltasol = 8.0\times 10^{-5}~{\rm eV^2} \,,
&
\sin^2\theta_{21}=0.31 \,, \\ [0.25cm]
|\deltaatm| = 2.2\times 10^{-3}~{\rm eV^2} \,,
&
\sin^22\theta_{23} = 1.0\,;\\[0.30cm]
\deltasol = (7.3 - 8.5) \times 10^{-5}~{\rm eV^2} \,,
&
\sin^2 \theta_{12} = (0.26 - 0.36)\;,\\ [0.25cm]
|\deltaatm| = (1.7 - 2.9)\times 10^{-3}~{\rm eV^2} \,,
&
\sin^22\theta_{23} \geq 0.90 \,.
\end{array}
\eeq
%
A combined 3-$\nu$ oscillation analysis of the global data
gives~\cite{BCGPRKL2,TSchwSNOW06}
\beq
\sin^2\theta_{13} < 0.027~(0.044)
\quad\mbox{at}\quad 95\%~(99.73\%)~{\rm C.L.}
\label{th13}
\eeq
%
Using the recently
announced (but still unpublished) 
data from the MINOS experiment~\cite{MINOS} 
in the analysis leads to a somewhat different best fit
value and 95\% allowed range of $|\deltaatm|$
and to a somewhat more stringent limit on $\sin^2\theta_{13}$
\cite{TSchwSNOW06}:
$|\deltaatm| = 2.6\times 10^{-3}~{\rm eV^2}$, 
$|\deltaatm| = (2.2 - 3.0)\times 10^{-3}~{\rm eV^2}$, and
$\sin^2\theta_{13} < 0.025~(0.040)$ at~95\%~(99.73\%) C.L.

   In spite of the enormous progress made in establishing 
the existence of neutrino oscillations driven by non-zero 
neutrino masses and mixing and in determining the 
pattern of neutrino mixing and the values of 
the two neutrino mass squared differences,
our knowledge and understanding of neutrino mixing is rather 
limited at present (see, e.g., \cite{STPNu04} for a detailed discussion 
of the current status of our ignorance about neutrino mixing).
Future progress 
in the studies of neutrino mixing 
requires, in particular, the knowledge of the precise 
values of the parameters which drive the solar and 
the dominant atmospheric neutrino oscillations, 
$\Delta m^2_{21}$, $\sin^2\theta_{12}$, $\Delta m^2_{31}$
and $\sin^2\theta_{23}$ (see, e.g., \cite{Freund:2001ui}).
The high precision measurement of these parameters
is one of the main goals of the next generation of 
neutrino oscillation experiments.
In the present article we discuss the 
possibility of a high precision
measurement of the solar neutrino 
oscillation parameters $\Delta m^2_{21}$ and
$\sin^2\theta_{12}$ in an experiment
studying the oscillations of 
reactor anti-neutrinos $\bar{\nu}_e$ 
with  a ``large scale'' detector 
located in the Frejus underground laboratory in France. 

  The existing data allow a determination of 
$\Delta m^2_{21}$ and $\sin^2\theta_{12}$
at 3$\sigma$ with an  error of 
approximately 11\% and 25\%, 
respectively. These parameters can 
and will be measured with higher precision 
in the future. The data from phase-III of 
the SNO experiment~\cite{sno} using 
$^3$He proportional counters
for the neutral current rate measurement
could lead to a reduction of the error 
in $\sin^2\theta_{12}$ 
to 21\%~\cite{SKGdCP04,BCGPTH1204}. 
If instead of 766.3~t~yr one uses simulated 3~kt~yr
KamLAND data in the same global solar and reactor neutrino 
data analysis, the 3$\sigma$ errors in 
$\Delta m^2_{21}$ and $\sin^2\theta_{12}$ 
diminish to 7\% and 18\%~\cite{BCGPTH1204}. 
The most precise measurement of 
$\Delta m^2_{21}$, discussed so far in the literature,
could be achieved \cite{SKGdCP04} using 
Super-Kamiokande doped with 0.1\% of Gadolinium 
(SK-Gd) for detection of reactor 
$\bar{\nu}_e$ \cite{SKGdBV04}:
the SK detector gets the same flux of reactor
$\bar{\nu}_e$ as KamLAND and
after 3 years of 
data-taking, $\Delta m^2_{21}$
could be determined with  
an error of 3.5\% at 3$\sigma$ 
\cite{SKGdCP04}. 
A dedicated reactor  
$\bar{\nu}_e$ experiment with a 
baseline $L\sim 60$~km, tuned to the minimum of the
$\bar{\nu}_e$ survival probability, 
could provide the most precise determination of $\sin^2\theta_{12}$
\cite{TH12}: with statistics of $\sim 60$ GW~kt~yr and a systematic error
of 2\% (5\%), $\sin^2\theta_{12}$ could be measured with an accuracy of
6\% (9\%) at 3$\sigma$ \cite{BCGPTH1204}.  The inclusion of the
uncertainty in $\theta_{13}$ ($\sin^2\theta_{13}<$0.05) in the
analyzes increases the quoted errors by (1--3)\% to approximately 9\%
(12\%) \cite{BCGPTH1204}.  
The improved determination of $\dms$ and
$\theta_{12}$ with KamLAND or dedicated post-KamLAND reactor neutrino
experiments has been studied previously also in
Refs.~\cite{Bandyopadhyay:2003ks,Schonert:2002ep,Bouchiat:2003jj,Minakata:2004jt,Kopp:2006mw},
whereas the potential improvements of the precision on these
parameters from future solar neutrino experiments has been
investigated, e.g., in
Refs.~\cite{BCGPTH1204,TH12,Bahcall:2003ce,Kopylov:2003sf}.

   MEMPHYS (MEgaton Mass PHYSics)~\cite{memphys} is a project for a
mega ton scale water \v{C}erenkov detector located in the Frejus
underground laboratory at the border of France and Italy.  It is
similar to the UNO~\cite{UNO} project in the US and the future
Hyper-Kamiokande~\cite{Nakamura:2003hk} detector in Japan. Such
detectors allow for a broad range of physics studies like nucleon
decay, long-baseline accelerator neutrino oscillations, super nova
neutrino detection, and oscillations of solar and atmospheric
neutrinos. The MEMPHYS detector is considered as a far detector for
neutrino beams produced at CERN located at a distance of 130~km from
Frejus (see, e.g., Ref.~\cite{Campagne:2006yx}). A recent civil
engineering pre-study indicates that MEMPHYS could be built with
existing techniques as a modular detector consisting of three (up to
five) modules (shafts), each having a fiducial mass of approximately
147~kt.

  In the present paper we consider the possibility that 
the water in one module of MEMPHYS is doped with 0.1\% of Gadolinium 
(MEMPHYS-Gd), as it has been proposed originally for 
Super-K~\cite{SKGdBV04}. This allows a very efficient
detection of electron anti-neutrinos through the reaction $\bar\nu_e +
p \to e^+ + n$ since the neutron can be tagged due to the high
absorption cross section on Gadolinium. One module 
of MEMPHYS (147~kt) is about 6.5 times bigger than Super-K (22.5~kt), which
increases correspondingly the potential for the various physics
applications, such as detection of relic or 
galactic super nova neutrinos, see Ref.~\cite{SKGdBV04}.
Here we explore the possibility of a precision measurement of the
solar neutrino oscillation parameters $\dms$ and $\theta_{12}$ by 
studying the oscillations of electron anti-neutrinos
emitted by the nuclear reactors located
in the ``neighborhood'' of the Gd-doped MEMPHYS detector.
We will compare, in particular, the 
precision on $\dms$ and $\theta_{12}$ which can be reached
with the MEMPHYS-Gd detector with that
obtainable with the Gd-doped Super-K detector (SK-Gd).
The latter has been studied in detail in Ref.~\cite{SKGdCP04}.

  The water \v{C}erenkov detectors typically do not have
very good energy resolution, which 
is compensated to certain extent by their large mass.
In what regards the energy resolution, 
the scintillator detectors such as KamLAND, perform
significantly better. The LENA (Low Energy Neutrino Astronomy)
detector~\cite{lena} is a project for a large ($\sim 50$~kt) liquid
scintillator detector, to be used for studies of relic and galactic
super nova neutrinos, solar neutrinos, geo-neutrinos, or proton
decay. Since neutrinos from nuclear reactors constitute a background for
the indicated measurements, some of the considered sites for LENA 
are rather far away from high concentrations of nuclear power plants.
We consider in the following the possibility to place a
LENA type detector in the Frejus laboratory (with many 
reactors relatively close by) and to use 
it for a high precision measurement of the oscillations
of reactor anti-neutrinos.

%
\section{Reactor Neutrino Measurements with MEMPHYS-Gd and LENA Detectors 
at Frejus}
%
\indent

  To calculate the flux of anti-neutrinos from reactors at a given
position on the Earth, public available information on the 
nuclear power plants can be used~\cite{reactor-info}. 
A list of, and relevant data on reactors  
compiled from such sources has been 
kindly provided to us~\cite{priv-com} for this study.
To compare the reactor neutrino fluxes at Frejus with those at Kamioka, Japan,
we consider first the effective reactor power at the detector, 
which is directly related to the total reactor $\bar{\nu}_e$ flux 
reaching the detector:
\begin{equation}\label{eq:Peff}
W_\mathrm{eff} = \sum_i \frac{W_i^\mathrm{th}}{4\pi L_i^2} \,, 
\end{equation}
%
where $W_i^\mathrm{th}$ 
is the thermal power of the $i$'th reactor, $L_i$ is the distance
to the detector, and the the sum runs over all contributing reactors.
We find that for Kamioka 
$W_\mathrm{eff}^\mathrm{Kamioka} \approx 3.0\,\rm MW \, km^{-2}/(4\pi)$,
whereas at Frejus the reactor $\bar{\nu}_e$ flux is slightly higher with
$W_\mathrm{eff}^\mathrm{Frejus} \approx 3.4\,\rm MW\, km^{-2}/(4\pi)$. 

\begin{figure}[!t]
\centering 
\includegraphics[width=0.57\textwidth]{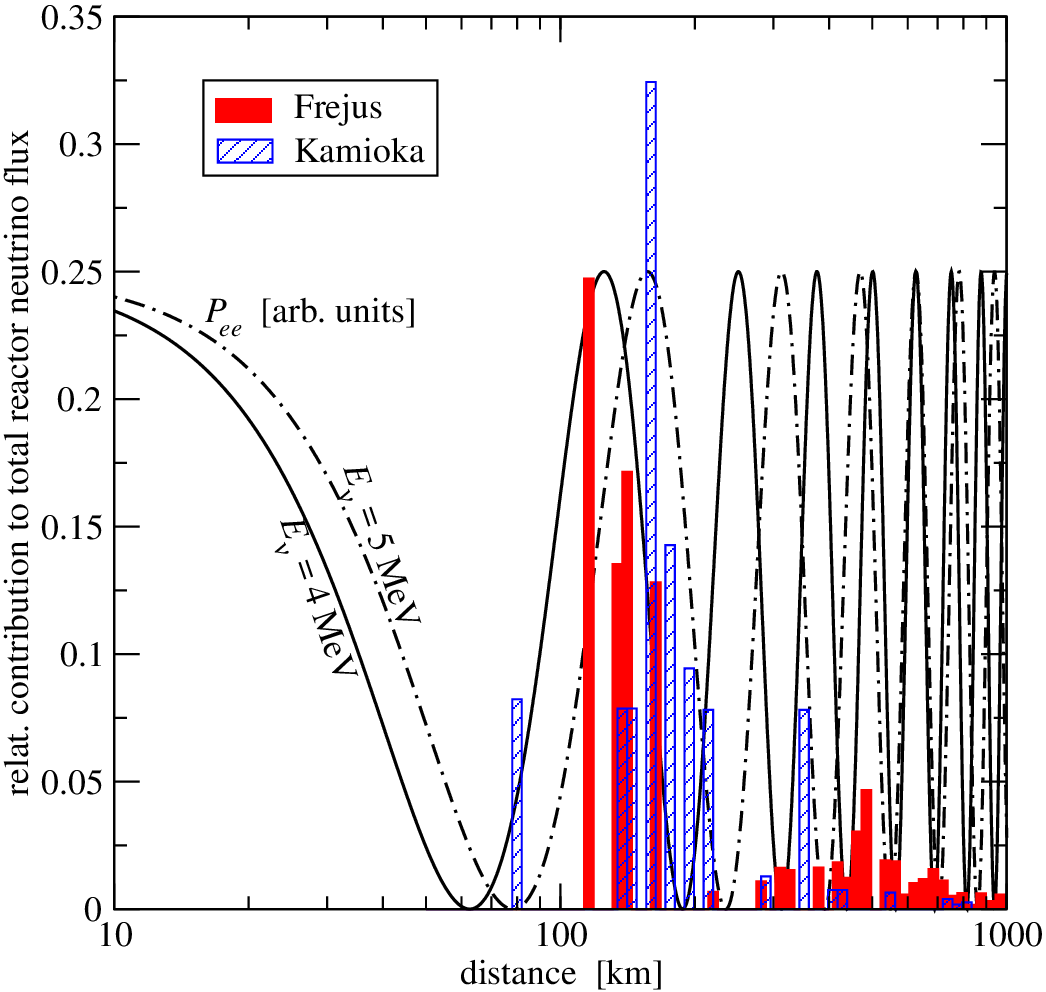}
  \mycaption{Relative contribution of different reactors to the
  total reactor neutrino flux at
  Frejus and Kamioka as a function of the distance to the reactor. 
  Also shown are the $\bar\nu_e$ survival probabilities 
  for $E_\nu = 4$~MeV and 5~MeV in arbitrary units.}
\label{fig:L-distr}
\end{figure}

The average distance traveled by reactor anti-neutrinos, $\langle
L\rangle = (\sum_i W_i^\mathrm{th}/L_i) /(4\pi W_\mathrm{eff})$, is
188~km for Kamioka and 299~km for Frejus.  Note that for $\dms =
8\times 10^{-5}$~eV$^2$ and a neutrino energy $E_{\nu} \sim 4$ MeV
(corresponding to the maximum of the event rate in the absence of
oscillations, see, e.g., the first article quoted in
Ref. \cite{Schonert:2002ep}) the first oscillation minimum of the
reactor $\bar{\nu}_e$ survival probability is at approximately
60~km. Therefore, the $\langle L\rangle$ for Frejus seems to be rather
large for an optimal measurement of the oscillation parameters.
However, the average distance can be misleading, and one should look
at the $L$-distribution of the reactor $\bar{\nu}_e$ flux.  In
Fig.~\ref{fig:L-distr} we show the relative contribution of different
reactors to the total reactor $\bar{\nu}_e$ flux at Frejus and Kamioka
as a function of the baseline. It turns out that 67\% of the total
flux at Frejus originates from four reactors along the Rhone river
located within a distance of 160~km from Frejus: Bugey at 115~km
(25\%), Saint Alban at 133~km (13\%), Cruas at 142~km (16\%), and
Tricastin at 160~km (13\%) \footnote{To use these four reactors for a
measurement of $\theta_{12}$ and $\dms$ has been considered previously
in Ref.~\cite{Bouchiat:2003jj}.}.  Approximately 31\% of the total
flux comes from reactors distributed between 300~km and 1000~km. In
our analysis we include 56 reactors located at a distance $L <
1000$~km, while the contributions of reactors at $L > 1000$~km from
all around the world are summed to one ``effective reactor'' at
2500~km giving 2\% of the total reactor $\bar{\nu}_e$ flux at Frejus.

  The comparison of the $L$-distribution with the $\bar{\nu}_e$ 
survival probability in Fig.~\ref{fig:L-distr} 
shows that the 4 reactors providing the dominant part of the
$\bar{\nu}_e$ flux at Frejus are  located at distances which permit 
a rather precise study of reactor $\bar{\nu}_e$ oscillations. 
For a $\bar{\nu}_e$ energy $E_{\nu} \sim (4 - 5)$~MeV
they are located between the first and the second survival
probability minima, and hence spectral information should
provide a powerful tool to measure the oscillation parameters. 
In the case of Kamioka the $L$-distribution is rather centered around
$\langle L\rangle \approx 190$~km. In Fig.~\ref{fig:L-distr} the
important contribution to the $\bar{\nu}_e$ flux from the Kashiwazaki
reactor complex located at approximately 160~km from Kamioka is
clearly visible. For $E_{\nu} \sim 5$~MeV this distance corresponds to
the first $\bar{\nu}_e$ survival probability maximum (see
Ref.~\cite{Bandyopadhyay:2003ks} for a detailed discussion).

  We calculate the observed prompt energy spectrum by
\begin{equation}\label{eq:event-spect}
\frac{dN}{dE_p} = \mathcal{N} \sum_i \frac{1}{4\pi L_i^2}
\int dE_\nu  \, \sigma(E_\nu) \, \phi_i(E_\nu)  \, P_{ee}(L_i, E_\nu) \,
R(E_p^\mathrm{tr}, E_p) \,,
\end{equation}
%
where the sum runs over the different reactors, $\sigma(E_\nu)$ is the
cross section of the detection reaction $\bar\nu_e + p \to e^+ + n$,
$P_{ee}$ is the $\bar\nu_e$ survival probability, and
$R(E_p^\mathrm{tr}, E_p)$ is the resolution function relating the
``true prompt energy'' $E_p^\mathrm{tr}$ to the prompt energy $E_p$
observed in the detector, where $E_p^\mathrm{tr}$ is determined by the
initial neutrino energy, $E_p^\mathrm{tr} = E_\nu - (m_n - m_p) + m_e
\cong E_{\nu} - 0.8$ MeV.  We work with the total prompt energy
visible in a scintillator detector (also if the actual detector
considered is water \v{C}erenkov) for the sake of comparison with
KamLAND. For $R(E_p^\mathrm{tr},E_p)$ we use a Gaussian resolution
function with mean $E_p^\mathrm{tr}$, and a width of $44\% / \sqrt{E_p
\,\rm[MeV]}$ for MEMPHYS-Gd/SK-Gd and $10\% / \sqrt{E_p \,\rm[MeV]}$
for LENA~\cite{lena}. The energy resolution for MEMPHYS-Gd and SK-Gd
is similar to the one reported by Super-K for the solar neutrino
analysis (see Fig.~15 of the second paper in Ref.~\cite{SKsolar}).

In Eq.~(\ref{eq:event-spect}), $\phi_i(E_\nu)$ denotes the
flux of $\bar\nu_e$ emitted by reactor $i$, which is given by
\begin{equation}\label{eq:flux}
\phi_i(E_\nu) = W^\mathrm{th}_i
\sum_\ell \frac{f_\ell}{E_\ell} \phi_\ell(E_\nu) \,,
\end{equation}
where $\ell =$~$^{235}$U, $^{238}$U, $^{239}$Pu, $^{241}$Pu, labels
the most important isotopes contributing to the $\bar\nu_e$ flux,
$f_\ell$ is the relative contribution of the isotope $\ell$ to the
total reactor power, and $E_\ell$ is the energy release per fission for
the isotope $\ell$. In Eq.~(\ref{eq:flux}), $\phi_\ell(E_\nu)$ is the
(energy differential) number of neutrinos emitted per fission by the
isotope $\ell$, and we adopt the parameterization for the
$\phi_\ell(E_\nu)$ from Ref.~\cite{Huber:2004xh}. For the $f_\ell$ we
take a typical isotope composition in a nuclear reactor
of~\cite{kamland1} $^{235}$U : $^{238}$U : $^{239}$Pu : $^{241}$Pu =
0.568 : 0.297 : 0.078 : 0.057, and we assume these ratios to be equal
for all reactors.

In the calculations we use the 3-neutrino oscillation
survival probability $P_{ee}$ which depends, in particular, on 
$\sin^2\theta_{13}$ (see, e.g., \cite{STPNu04}),
and take into account the (small) Earth matter effect.
In the case of absence of oscillations, $P_{ee} = 1$, the number of
events above a threshold $E_{\mathrm{thr}}$ is given by
$N_\mathrm{no\,osc} = \mathcal{N} W_\mathrm{eff} C$, where
$W_\mathrm{eff}$ has been defined in Eq.~(\ref{eq:Peff}) and $C$ is an
integral depending only on $E_\mathrm{thr}$. For MEMPHYS-Gd we use a
threshold for the prompt energy $E_\mathrm{thr} = 3.0$~MeV (which
corresponds to the value of 2.5~MeV for the positron energy given in
Ref.~\cite{SKGdBV04}), whereas for LENA we use $E_\mathrm{thr} =
2.6$~MeV to eliminate the background from geo-neutrinos, as in the
KamLAND oscillation analysis~\cite{kamland1,kamland}.
To determine the normalization constant $\mathcal{N}$ in
Eq.~(\ref{eq:event-spect}) we use the prediction for
$N_\mathrm{no\,osc}$ in KamLAND~\cite{kamland1}, 
and then we scale it for each experiment taking 
into account that $\mathcal{N}$ is
proportional to the measurement time and the number of free protons in
the detector, as well as the different values of 
$W_\mathrm{eff}$ and $E_\mathrm{thr}$.  
In Tab.~\ref{tab:exps} we summarize the most important characteristics
of the considered detectors as simulated in our analysis, and we give
the expected number of events in case of no oscillations.

\begin{table}
\centering
\catcode`?=\active \def?{\hphantom{0}}
\begin{tabular}{lcccccc}
\hline\hline
experiment & fid.\ mass & free protons & $E_\mathrm{thr}$ & events/yr
& energy resol. \\
\hline
MEMPHYS-Gd & 147.0? kt & $9.8\times 10^{33}$ & 3.0 MeV & 59 980 
        & $44\% / \sqrt{E_p\,\rm[MeV]}$\\
LENA    & ?44.0? kt & $2.3\times 10^{33}$ & 2.6 MeV & 16 670 
        & $10\% / \sqrt{E_p\,\rm[MeV]}$\\
SK-Gd   & ?22.5? kt & $1.5\times 10^{33}$ & 3.0 MeV & ?8 000 
        & $44\% / \sqrt{E_p\,\rm[MeV]}$\\
KamLAND & ??0.41 kt & $3.5\times 10^{31}$ & 2.6 MeV & ?? 216 
        & $?7.5\% / \sqrt{E_p\,\rm[MeV]}$\\
\hline\hline
\end{tabular}
  \mycaption{Summary of the input characteristics  
  of the detectors MEMPHYS-Gd,
  LENA, and SK-Gd, used in our analysis. 
  For comparison we show also the corresponding
  values for KamLAND~\cite{kamland1}. The number of events/yr is
  calculated for no oscillations and using the reactor flux at Frejus
  for MEMPHYS-Gd and LENA, and at Kamioka for SK-Gd.}
\label{tab:exps}
\end{table}

\begin{figure}[!t]
\centering 
\includegraphics[width=0.6\textwidth]{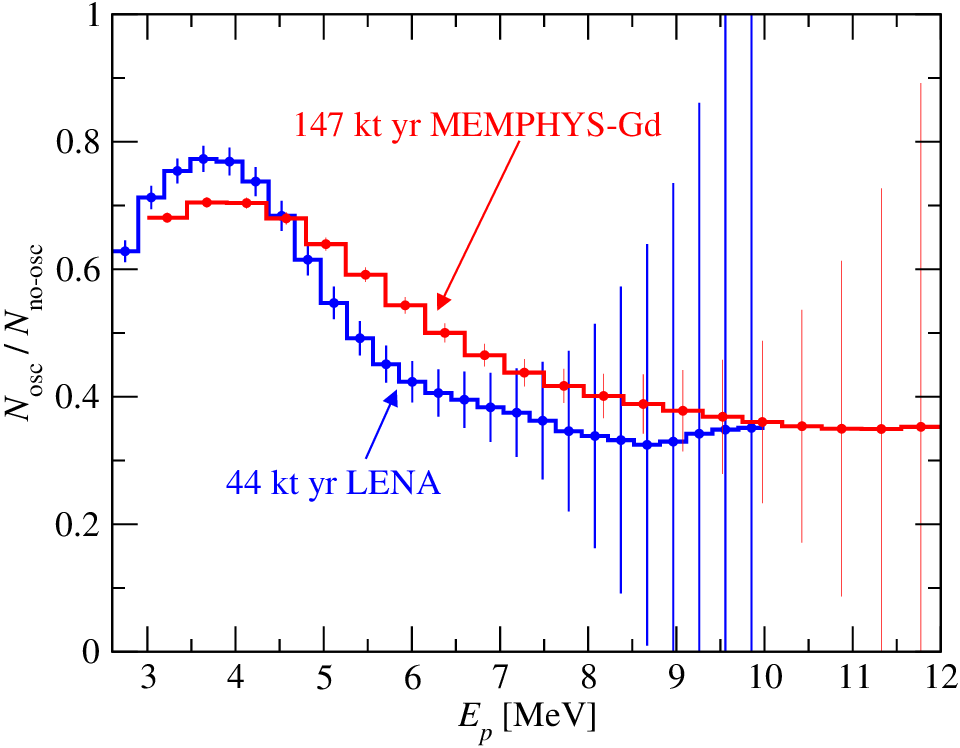}
  \mycaption{The ratio of the event spectra in positron energy 
  in the case of oscillations with $\dms = 7.9\times 10^{-5}$~eV$^2$ and
  $\sin^2\theta_{12} = 0.30$ and in the absence of oscillations, 
  determined using one year data of MEMPHYS-Gd and LENA located at Frejus. 
  The error bars correspond to $1\sigma$ statistical error.}
\label{fig:histo}
\end{figure}

  To test the sensitivity of the experiments we divide the prompt energy 
spectrum in Eq.~(\ref{eq:event-spect}) into 20 bins between 3~MeV
and 12~MeV for MEMPHYS-Gd and SK-Gd, and into 25 bins between 2.6~MeV and
10~MeV for LENA
\footnote{For low statistics data samples such as the present KamLAND
one, a likelihood analysis~\cite{Schwetz:2003se} or equal bins in
$1/E_p$~\cite{TSchwSNOW06} allows to extract an optimum of
information. In the cases under study, however, the numbers of events
are relatively high. Therefore, we simply take equal bins in $E_p$,
with a bin size sufficiently smaller than the energy resolution to
make sure that no information is lost due to the binning.}.  
The data is simulated using as ``true values'' $\dms = 7.9\times
10^{-5}$~eV$^2$ and $\sin^2\theta_{12} = 0.30$. The latter correspond
to the present best fit point obtained in a global neutrino
oscillation analysis~\cite{TSchwSNOW06}. Then a $\chi^2$-analysis is
performed to determine the allowed regions and the precision with
which these parameters can be determined from the simulated data.
In Fig.~\ref{fig:histo} we show the ratio of the number of events
calculated by taking into account $\bar{\nu}_e$ oscillations with
parameters indicated above to the number of events in the absence of
oscillations, binned in prompt energy.  The error bars correspond to
$1\sigma$ statistical errors for one year of (simulated)
MEMPHYS-Gd and LENA data. The large number of events leads to a very
precise measurement of the energy spectrum. The oscillatory signal in
the spectrum is less pronounced in the MEMPHYS-Gd spectral ``data''
than in the analogous LENA ``data'' due to the worse energy resolution
of the water \v{C}erenkov detector.  Nevertheless, as a consequence of
the relatively high statistics, a clear signal of spectral distortion
can still be observed with MEMPHYS-Gd.  In the case of LENA spectral
``data'', an event maximum is clearly visible at $E_p = (3.5-4.0)$
MeV, which originates from the first oscillation maximum of the
survival probability at $L \cong 160$ km (see
Fig.~\ref{fig:L-distr}). In both cases the spectral information is
crucial for the precise determination of the oscillation parameters.

\begin{table}
\centering
\catcode`?=\active \def?{\hphantom{0}}
\begin{tabular}{lr}
\hline\hline
systematic & value \\
\hline
overall normalization (fully correlated)      & 5\%   \\
thermal power of each reactor (uncorrelated)  & 2\%   \\
energy scale uncertainty                      & 0.5\% \\
prompt energy spectrum tilt                   & 2\%   \\
reactor neutrino flux            & Ref.~\cite{Huber:2004xh}\\
\hline\hline
\end{tabular}
  \mycaption{Systematical uncertainties and the default values
  adopted in our analysis.}
\label{tab:systematics}
\end{table}

In the statistical analysis we take into account various systematical
uncertainties as listed in Tab.~\ref{tab:systematics}. We include a
5\% error on the overall normalization (e.g., from the uncertainty on
the fiducial mass), a 2\% uncertainty on the thermal power of each
reactor (uncorrelated between the reactors), and the uncertainty on
the anti-neutrino spectra $\phi_\ell(E_\nu)$ (normalization and shape)
according to Ref.~\cite{Huber:2004xh}. We take into account an
uncertainty of 0.5\% in the energy scale calibration of the
detector. This value is motivated by the numbers given for the Super-K
solar neutrino analysis (see second reference in~\cite{SKsolar}) and
for the Double-Chooz reactor
experiment~\cite{Double-Chooz}. Furthermore, we include a linear tilt
in the prompt energy spectrum of 2\%, i.e., we allow the event number
in the highest energy bin to shift by 2\% with respect to the event
number in the lowest energy bin with a linear interpolation for the
intermediate bins. In the following section we will discuss in some
detail how much our numerical results depend on the values adopted for
the systematic uncertainties.
In addition to these systematics we have tested also the effect of an
uncertainty on the isotope compositions $f_\ell$ defined in
Eq.~(\ref{eq:flux}). We have performed an analysis allowing the
$f_\ell$ to vary independent for each reactor within 5\%, and found
that the impact on the sensitivity to the neutrino oscillation
parameters is negligible. Therefore, we keep the $f_\ell$ fixed in our
standard analysis which significantly reduces the calculation time.

%
\section{Precision of the Determination of 
Neutrino Oscillation Parameters}
%
\indent

\begin{figure}[!t]
\centering 
\includegraphics[width=0.7\textwidth]{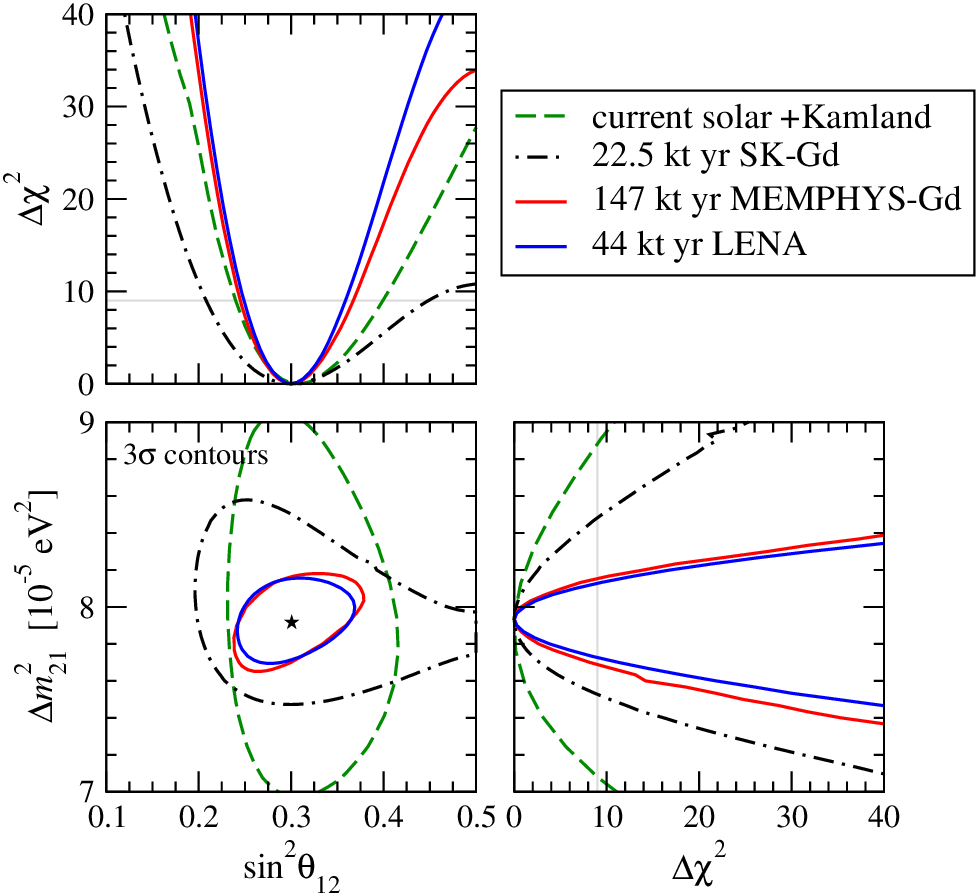}
  \mycaption{The accuracy of the determination of $\dms$ and
  $\sin^2\theta_{12}$, which can be obtained using one year of data
  from MEMPHYS-Gd and LENA at Frejus, and from SK-Gd at Kamioka,
  compared to the current precision from solar neutrino and KamLAND
  data. We show the allowed regions at $3\sigma$ (2 d.o.f.) in the
  $\dms-\sin^2\theta_{12}$ plane, as well as the projections of the
  $\chi^2$ for each parameter.}
\label{fig:sensitivities}
\end{figure}

 In this Section we present results on the precision which
can be reached in the measurement of 
$\dms$ and $\sin^2\theta_{12}$ in the MEMPHYS-Gd and LENA 
experiments. Our findings are summarized in 
Fig.~\ref{fig:sensitivities} and
Tab.~\ref{tab:spread}, where we compare the results 
which can be obtained using one year of
data from MEMPHYS-Gd, LENA, and SK-Gd, with 
the present constraints from the global solar and 
KamLAND data~\cite{BCGPRKL2,TSchwSNOW06}. In the table we
give the relative uncertainty at $3\sigma$ defined as 
\begin{equation}\label{eq:spread}
\mbox{spread}(x) = 
\frac{x^\mathrm{upper} - x^\mathrm{lower}}
{x^\mathrm{upper} + x^\mathrm{lower}}\,,
\end{equation}
%
where $x^\mathrm{upper} (x^\mathrm{lower})$ is the upper (lower) bound of
the quantity $x$ at $3\sigma$.

\begin{table}[!t]
\centering
\catcode`?=\active \def?{\hphantom{0}}
\begin{tabular}{l@{\qquad}cccc}
\hline\hline
      & \multicolumn{2}{c}{spread($\dms$)} & 
        \multicolumn{2}{c}{spread($\sin^2\theta_{12}$)} \\
time       & 1 yr   & 7 yr  & 1 yr   & 7 yr   \\
\hline
SK-Gd      & 6.0\%  & 2.8\% & 36.6\% & 18.6\% \\
MEMPHYS-Gd & 2.9\%  & 1.4\% & 20.0\% & 13.2\% \\
LENA       & 2.5\%  & 1.2\% & 18.0\% & ?9.8\% \\
\hline
solar + KamLAND  & \multicolumn{2}{c}{11.3\%} & 
                   \multicolumn{2}{c}{24.9\%} \\
\hline\hline
\end{tabular}
  \mycaption{The $3\sigma$ uncertainty as defined in Eq.~(\ref{eq:spread}) 
  in the determination of $\dms$ and $\sin^2\theta_{12}$
  in the experiments MEMPHYS-Gd and LENA at Frejus and SK-Gd at Kamioka 
  after 1 and 7 years of data taking.   
  For comparison we show also the current uncertainties in
  $\dms$ and $\sin^2\theta_{12}$~\cite{TSchwSNOW06}.}
\label{tab:spread}
\end{table}

\begin{figure}[!t]
\centering 
\includegraphics[width=0.7\textwidth]{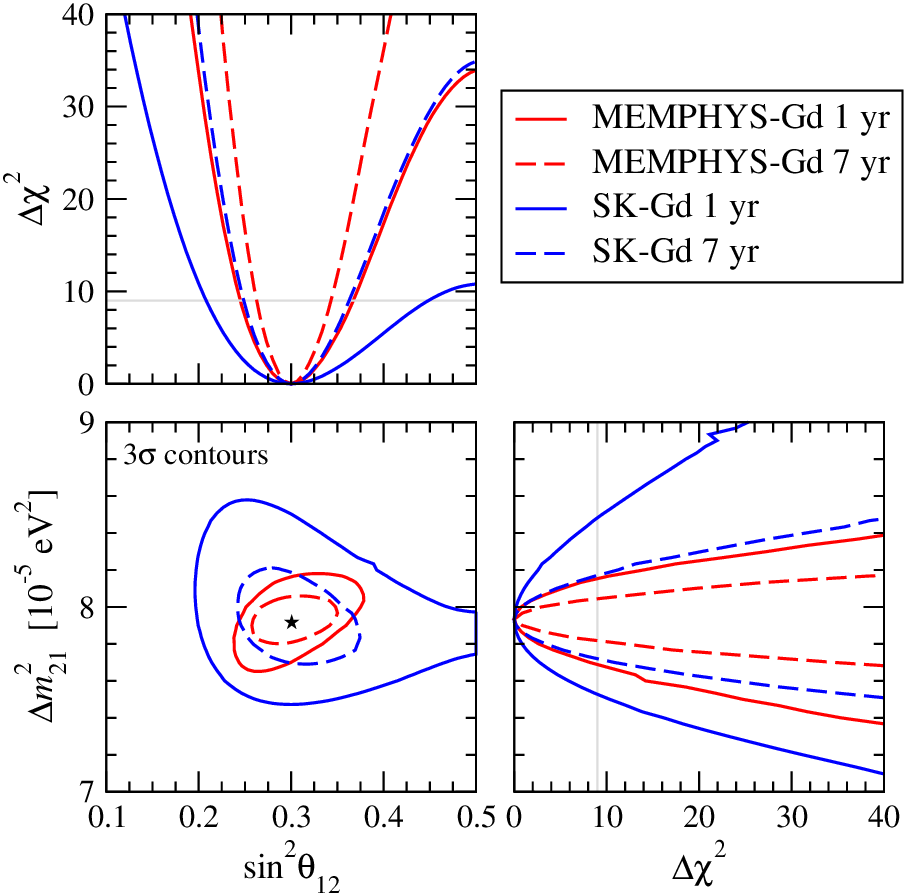}
  \mycaption{The allowed regions in the $\dms-\sin^2\theta_{12}$
   plane, obtained at $3\sigma$ (2 d.o.f.)  from 147~kt~yr and
   $147\times 7$~kt~yr of MEMPHYS-Gd, and from 22.5~kt~yr and
   $22.5\times 7 = 157.5$~kt~yr of SK-Gd data.  The projections of the
   $\chi^2$ for each parameter are also shown.}
\label{fig:mem-vs-sk}
\end{figure}

  We find that the water \v{C}erenkov detector MEMPHYS-Gd and the
scintillator detector LENA can provide very similar high precision
determinations of $\dms$ and $\sin^2\theta_{12}$. The lower mass of
LENA is compensated by its better energy resolution. Already with one
year of data, an uncertainty smaller than 3\% at $3\sigma$ can be
obtained in the determination of $\dms$, while $\sin^2\theta_{12}$ can
be determined with an error of about 20\% at $3\sigma$.  This
precision is approximately by a factor two better than the precision
that can be reached with one year of data from SK-Gd.
The better precisions which can be obtained with the
MEMPHYS-Gd detector compared to those that can be 
obtained with the SK-Gd detector are a consequence
of the larger fiducial mass of MEMPHYS-Gd. 
As follows from Tab.~\ref{tab:spread} and Fig.~\ref{fig:mem-vs-sk}, 
1~year of data from MEMPHYS-Gd and 7~years of data from SK-Gd
(yielding approximately the same numbers of events in the two
detectors, see Tab.~\ref{tab:exps})
allow a determination of $\dms$ and $\sin^2\theta_{12}$
with similar precisions. This shows, in particular, 
that both locations, Frejus and Kamioka, are very similar 
in what regards the power of the ``surrounding'' reactors 
and their distance distribution. 
Thus, the two  locations are equally suitable 
for the high precision measurements under discussion.

  Ultimately, 7 years of data from MEMPHYS-Gd (LENA) would allow a
determination of $\dms$ and $\sin^2\theta_{12}$ with uncertainties of
approximately 1.4\% (1.2\%) and 13\% (10\%) at 3$\sigma$,
respectively. This precision is comparable to the precision which can
be reached in the determination of the atmospheric neutrino
oscillation parameters $\dma$ and $\sin^2\theta_{23}$ by studying
$\nu_\mu$ disappearance in the superbeam experiments T2HK in Japan or
SPL from CERN to MEMPHYS (see e.g., Ref.~\cite{Campagne:2006yx} for a
recent analysis). Hence, the reactor measurement could complete the
program of the high precision determination of the parameters responsible
for the leading solar and atmospheric neutrino oscillations.

\begin{figure}[!t]
\centering 
\includegraphics[width=0.7\textwidth]{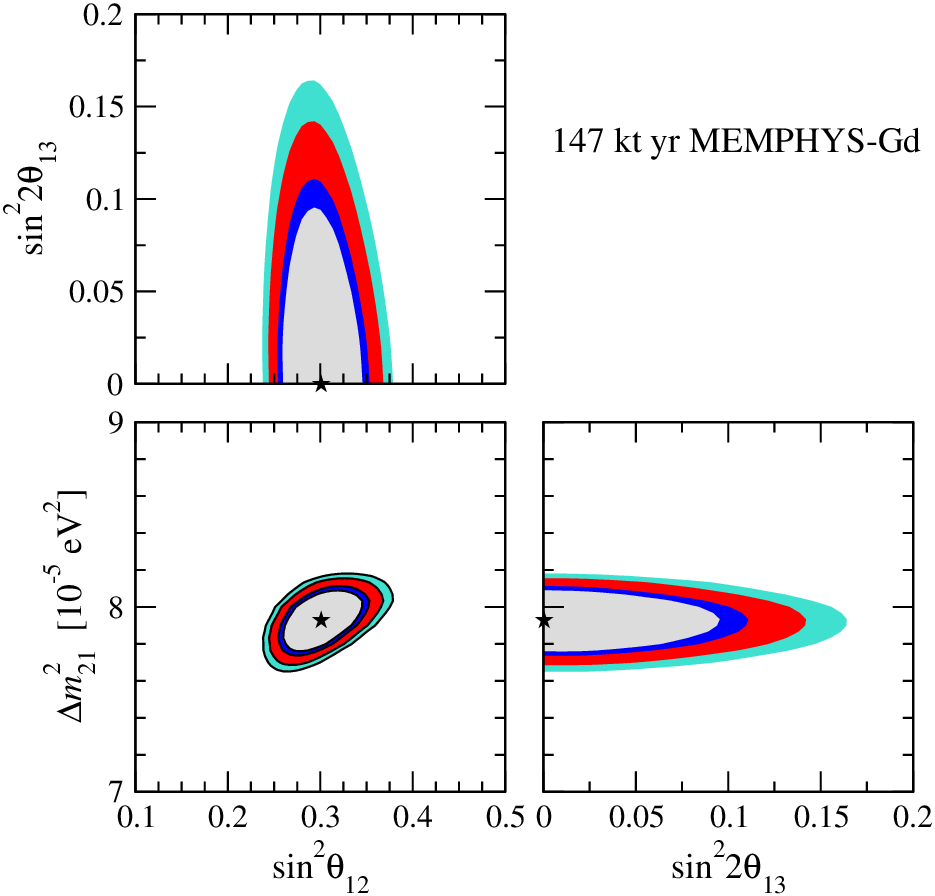}
  \mycaption{Allowed regions from a three-flavour neutrino oscillation
   analysis of 147~kt~yr data in MEMPHYS-Gd at 90\%, 95\%, 99\%, and
   99.73\%~CL (2~d.o.f.), projected onto the three different
   2-dimensional parameter planes. The data are simulated for
   $\theta_{13} = 0$. In the analysis $\theta_{13}$ was varied freely
   taking into account the constraint from the current global data as
   obtained in Ref.~\cite{TSchwSNOW06}. The allowed regions in the
   $\sin^2\theta_{12}-\dms$ plane from a two-flavour analysis with
   $\theta_{13}$ fixed to zero (indicated with black contours) are
   identical to those obtained in the three-flavour analysis.}
\label{fig:mem_3nu}
\end{figure}

We have also investigated whether the uncertainty in the knowledge of
$\theta_{13}$ can have any effect on the precision of $\dms$ and
$\theta_{12}$ determination in the experiments MEMPHYS-Gd and LENA
under discussion.  We show in Fig.~\ref{fig:mem_3nu} the results of a
three-flavour neutrino oscillation analysis of 1 year simulated data
in MEMPHYS-Gd, assuming that the true value of $\theta_{13}$ is zero.
In the analysis $\sin^2\theta_{13}$ was allowed to vary freely,
however the information available at present has been included in the
fit by adding the $\chi^2(\theta_{13})$ obtained from current global
data in Ref.~\cite{TSchwSNOW06}. In the panel for the
$\sin^2\theta_{12}-\dms$ projection we show also the allowed regions,
obtained in a two-flavour analysis, i.e., for $\theta_{13} = 0$, with
black contours. Indeed, the regions within the black contours are
practically identical to the shaded/colored regions corresponding to
the three-flavour case. Therefore we can conclude that the uncertainty
in the knowledge of $\theta_{13}$ does not affect the $\dms$ and
$\theta_{12}$ measurements. As is visible in Fig.~\ref{fig:mem_3nu},
there are no correlations of the leading parameters with
$\theta_{13}$, since $\dms$ and $\theta_{12}$ are determined primarily
from the data on the shape of the spectrum, which does not depend on
$\theta_{13}$.  We have checked that the above conclusions concerning
the $\theta_{13}$-effects hold also for LENA and SK-Gd detectors.

  Let us note that the sensitivity of MEMPHYS-Gd to $\theta_{13}$ on
its own is rather poor. The constraint on $\sin^2\theta_{13}$
appearing in Fig.~\ref{fig:mem_3nu} corresponds just to the bound from
present global data, which is included in the analysis. Hence, the
sensitivity of MEMPHYS-Gd is worse than the present bound. This is a
consequence of the fact that a non-zero $\theta_{13}$ basically leads
to a rather small (energy-independent) suppression of the total
$\bar{\nu}_e$ flux, which is unobservable due to the relatively large
uncertainties in the overall normalization.

\begin{figure}[!t]
\centering 
\includegraphics[width=0.8\textwidth]{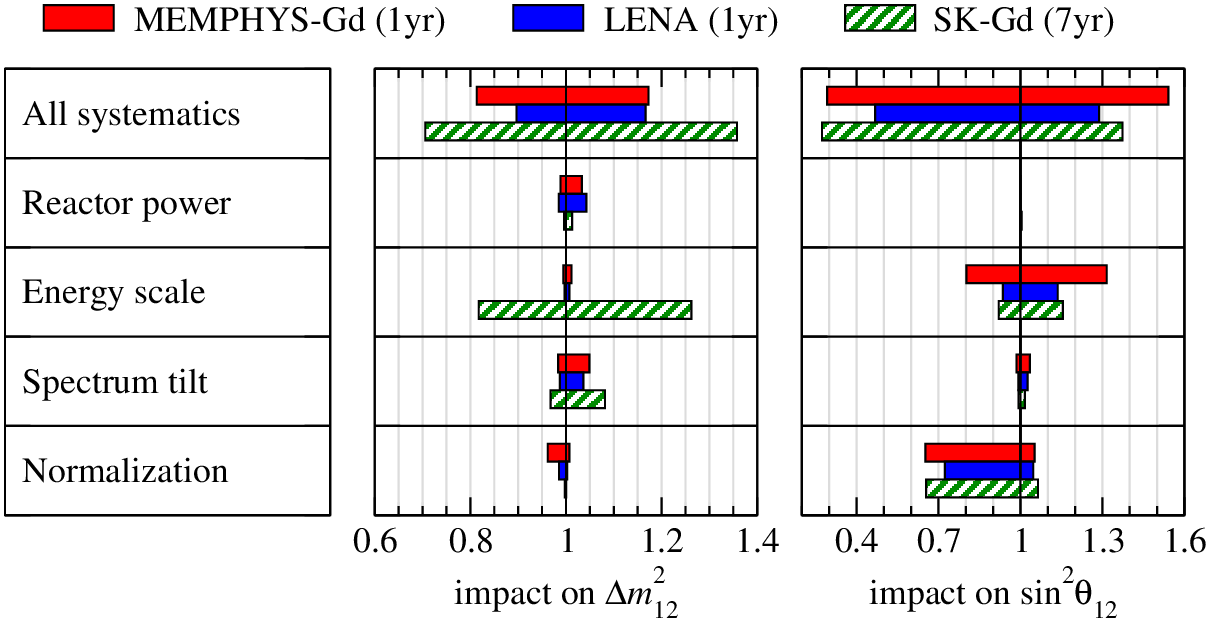}
  \mycaption{Impact of systematical uncertainties. We show the
  relative change in the $3\sigma$ intervals of $\dms$ and
  $\sin^2\theta_{12}$ obtained by switching off (left edges of the
  bars) and doubling (right edges) all systematics simultaneously, the
  uncertainties on the thermal power of the 4 most important reactors
  (``Reactor power''), the energy scale uncertainty, the prompt energy
  spectrum tilt, and the overall normalization error.}
\label{fig:systematics}
\end{figure}

Since we are discussing here very high statistics measurements, a
careful investigation of the impact of systematical uncertainties is
necessary. In Fig.~\ref{fig:systematics} we show how our results for
the accuracies of the oscillation parameters depend on the assumptions
adopted for the systematical errors. In particular, we consider the
impact of uncertainties on the thermal power of the 4 most important
reactors, the energy scale uncertainty, the prompt energy spectrum
tilt, the overall normalization error, as well as all systematical
errors in total. To check the impact of these uncertainties we show in
Fig.~\ref{fig:systematics} the ratios $\delta_0/\delta_\mathrm{std}$
and $\delta_2/\delta_\mathrm{std}$, where $\delta_{0(2)}$ is the
$3\sigma$ range for $\dms$ or $\sin^2\theta_{12}$ if the systematical
error of interest is set to zero (is doubled with respect to its
standard value), and $\delta_\mathrm{std}$ is the $3\sigma$ range
using our standard values according to Tab.~\ref{tab:systematics}.

The left edges of the bars in the row denoted by ``All systematics''
in Fig.~\ref{fig:systematics} correspond to statistical errors
only. In this ideal case the $\dms$ accuracy is improved by about 20\%
for MEMPHYS-Gd, 10\% for LENA, and 30\% for SK-Gd with respect to our
standard choice for the systematics, whereas the precision on
$\sin^2\theta_{12}$ is improved for all experiments by more than a
factor of 2. 
For the $\dms$ measurement the individual systematics have only a
minor impact (with the exception of a $\sim 20\%$ effect of the energy
scale uncertainty in SK-Gd). For the measurement of $\sin^2\theta_{12}$
the overall normalization and the energy scale (especially for
MEMPHYS-Gd) are important. 
Note that the uncertainty on the thermal reactor power has a
negligible impact on the accuracies. Therefore, it seems not to be
possible to improve the precision on $\dms$ and $\theta_{12}$ by
installing near detectors close to the reactors dominating the
$\bar\nu_e$ flux.

In summary, systematic uncertainties are an important factor in the
experiments under consideration. Especially the determination of the
mixing angle depends on the values of systematic errors. The overall
effect emerges from an interplay of the various sources of
uncertainties included in our analysis, and therefore, to obtain an
improvement in the precision of the oscillation parameters several of
the systematic errors listed in Tab.~\ref{tab:systematics} should be
decreased.

\section{Conclusions}
\indent 

We have investigated the possibility of a high precision determination
of the solar neutrino oscillation parameters $\dms$ and $\theta_{12}$
in a long-baseline reactor neutrino experiment, located in the Frejus
underground laboratory. Approximately 67\% of the total reactor
$\bar{\nu}_e$ flux at Frejus originates from four nuclear power plants
in the Rhone valley, located at distances between 115~km and 160~km
from Frejus.  The indicated baselines are particularly suitable for
the study of the $\bar{\nu}_e$ oscillations driven by $\dms$---they
are similar to those exploited in the KamLAND experiment in Japan.
Approximately 31\% of the total flux $\bar{\nu}_e$ at Frejus comes
from reactors distributed between 300~km and 1000~km from Frejus.  In
our analysis we include 56 reactors located at a distance $L <
1000$~km, while the contributions of reactors at $L > 1000$~km from
all around the world are summed to one ``effective reactor'' at
2500~km giving 2\% of the total reactor $\bar{\nu}_e$ flux at Frejus.
The Frejus underground laboratory is under consideration as a possible
site for a mega ton scale water \v{C}erenkov detector MEMPHYS which,
among other physics applications, may serve as a far detector for a
neutrino beam produced at CERN. In the present article we have assumed
that the water of one module of MEMPHYS having a fiducial mass of
147~kt, is doped with 0.1\% Gadolinium (MEMPHYS-Gd), which will allow,
in principle, a high precision study of reactor $\bar{\nu}_e$
oscillations. As an alternative detector technology, we have
considered a 50~kt scale liquid scintillator detector, as discussed in
the LENA proposal, which can be viewed as a considerably larger
version of the present KamLAND or Borexino detectors.

The analysis performed by us shows that each of the two
detectors---MEMPHYS-Gd and LENA, if placed at Frejus, would allow a
very precise determination of the solar neutrino oscillation
parameters $\dms$ and $\sin^2\theta_{12}$: with one year of reactor
$\bar\nu_e$ data taken at Frejus (by any of the two detectors), the
3$\sigma$ uncertainties on $\dms$ and $\sin^2\theta_{12}$ can be
reduced respectively to less than 3\% and to approximately 20\%. The
Gadolinium doped Super-Kamiokande detector (SK-Gd) in Japan can reach
a similar precision if the SK/MEMPHYS fiducial mass ratio of 1 to 7 is
compensated by a longer SK-Gd data taking time.
Several years of reactor $\bar{\nu}_e$ data collected by 
MEMPHYS-Gd or LENA would allow a determination 
of $\dms$ and $\sin^2\theta_{12}$ with
uncertainties of approximately 1\% and 10\% at 3$\sigma$,
respectively. 
We have shown also that the uncertainty associated with the CHOOZ
mixing angle $\theta_{13}$ has practically no impact on the
measurements of the solar neutrino oscillation parameters in the
experiments discussed by us, and we have investigated in some detail
the effects of various systematical uncertainties on the precision of
the determination of $\dms$ and $\sin^2\theta_{12}$ in these
experiments.

The accuracies on the solar oscillation parameters, which can be
obtained in the high statistics experiments considered here are
comparable to those that can be reached for the atmospheric neutrino
oscillation parameters $\dma$ and $\sin^2\theta_{23}$ in future
long-baseline superbeam experiments like T2HK in Japan or SPL from
CERN to MEMPHYS. Hence, such reactor measurements would complete the
program of the high precision determination of the leading neutrino
oscillation parameters.

\vspace*{5mm}{\bf Acknowledgments.}
We are grateful to Alessandra Tonazzo and Michael Wurm for providing
us information on nuclear power plants, and we thank Jean-Eric
Campagne and Michael Wurm for useful communications on the MEMPHYS and
LENA detectors.
This work was supported in part by the Italian MIUR and INFN under the
programs ``Fisica Astroparticellare'' (S.T.P.). The work of T.S.\ is
supported by a ``Marie Curie Intra-European Fellowship within the 6th
European Community Framework Program.''

\end{document}